\documentclass[journal]{IEEEtran}
\makeatletter\if@twocolumn\PassOptionsToPackage{switch}{lineno}\else\fi\makeatother

\usepackage{graphicx}
\usepackage[T1]{fontenc}
\ifCLASSINFOpdf
\else
\fi
\usepackage{amsmath}
\usepackage[cmintegrals]{newtxmath}
\usepackage{url,multirow,morefloats,floatflt,cancel,tfrupee}
\makeatletter

\AtBeginDocument{\@ifpackageloaded{textcomp}{}{\usepackage{textcomp}}}
\makeatother
\usepackage{colortbl}
\usepackage{xcolor}
\usepackage{pifont}
\usepackage[nointegrals]{wasysym}
\makeatletter

\def\mcWidth#1{\csname TY@F#1\endcsname+\tabcolsep}

\def\cAlignHack{\rightskip\@flushglue\leftskip\@flushglue\parindent\z@\parfillskip\z@skip}
\def\rAlignHack{\rightskip\z@skip\leftskip\@flushglue \parindent\z@\parfillskip\z@skip}

\usepackage{ifxetex}
\ifxetex\else\if@twocolumn\usepackage{dblfloatfix}\fi\fi

\AtBeginDocument{
\expandafter\ifx\csname eqalign\endcsname\relax
\def\eqalign#1{\null\vcenter{\def\\{\cr}\openup\jot\m@th
  \ialign{\strut$\displaystyle{##}$\hfil&$\displaystyle{{}##}$\hfil
      \crcr#1\crcr}}\,}
\fi
}

\AtBeginDocument{%
  \@ifpackageloaded{endfloat}%
   {\renewcommand\efloat@iwrite[1]{\immediate\expandafter\protected@write\csname efloat@post#1\endcsname{}}}{}%
}%

\def\BreakURLText#1{\@tfor\brk@tempa:=#1\do{\brk@tempa\hskip0pt}}
\let\lt=<
\let\gt=>
\def\processVert{\ifmmode|\else\textbar\fi}

\@ifundefined{subparagraph}{
\def\subparagraph{\@startsection{paragraph}{5}{2\parindent}{0ex plus 0.1ex minus 0.1ex}%
{0ex}{\normalfont\small\itshape}}%
}{}

\newcommand\role[1]{\unskip}
\newcommand\aucollab[1]{\unskip}
  
\@ifundefined{tsGraphicsScaleX}{\gdef\tsGraphicsScaleX{1}}{}
\@ifundefined{tsGraphicsScaleY}{\gdef\tsGraphicsScaleY{.9}}{}
\def\checkGraphicsWidth{\ifdim\Gin@nat@width>\linewidth
	\tsGraphicsScaleX\linewidth\else\Gin@nat@width\fi}

\def\checkGraphicsHeight{\ifdim\Gin@nat@height>.9\textheight
	\tsGraphicsScaleY\textheight\else\Gin@nat@height\fi}

\def\fixFloatSize#1{}%\@ifundefined{processdelayedfloats}{\setbox0=\hbox{\includegraphics{#1}}\ifnum\wd0<\columnwidth\relax\renewenvironment{figure*}{\begin{figure}}{\end{figure}}\fi}{}}
\let\ts@includegraphics\includegraphics

\def\inlinegraphic[#1]#2{{\edef\@tempa{#1}\edef\baseline@shift{\ifx\@tempa\@empty0\else#1\fi}\edef\tempZ{\the\numexpr(\numexpr(\baseline@shift*\f@size/100))}\protect\raisebox{\tempZ pt}{\ts@includegraphics{#2}}}}

\AtBeginDocument{\def\includegraphics{\@ifnextchar[{\ts@includegraphics}{\ts@includegraphics[width=\checkGraphicsWidth,height=\checkGraphicsHeight,keepaspectratio]}}}

\DeclareMathAlphabet{\mathpzc}{OT1}{pzc}{m}{it}

\def\URL#1#2{\@ifundefined{href}{#2}{\href{#1}{#2}}}

\def\UrlOrds{\do\*\do\-\do\~\do\'\do\"\do\-}%
\g@addto@macro{\UrlBreaks}{\UrlOrds}

\@ifundefined{quoteAttrib}
	{}
	{}

\@ifundefined{titlequoteAttrib}
	{}{}

\newenvironment{title-quote}
	{\list{}{\fontsize{10pt}{12pt}\selectfont\leftmargin.5in\itshape\rightmargin\leftmargin}%
  \item\relax}
  {\endlist}

\makeatother

\makeatletter
\AtBeginDocument{\@ifpackageloaded{longtable}{%
\def\LT@makecaption#1#2#3{%
  \LT@mcol\LT@cols c{\hbox to\z@{\hss\parbox[t]\LTcapwidth{%
    \sbox\@tempboxa{#1{#2: } #3}%
    \ifdim\wd\@tempboxa>\hsize
      #1{#2: }\textsc{#3}%
    \else
      \hbox to\hsize{\hfil\box\@tempboxa\hfil}%
    \fi
    \endgraf\vskip\baselineskip}%
  \hss}}}
}{}}
\makeatother

   \makeatletter
  \def\fig@textbf{\textbf}
   \AtBeginDocument{\renewcommand\floatc@plain[2]{\setbox\@tempboxa\hbox{{\footnotesize#1.}\footnotesize\hskip.5em#2}%
    \ifdim\wd\@tempboxa>\hsize {\fig@textbf{\footnotesize#1.}}\footnotesize\hskip.5em#2\par
        \else\hbox to\hsize{\hfil\box\@tempboxa\hfil}\fi}}
    \makeatother
  
\usepackage{float}

\begin{document}


        % \title{Learning to Model Human Hearing Perception Using Neural Loss Functions}
      \title{Learning to Model Aspects of Auditory Perception Using Neural Loss Functions}
\author{Prateek~Verma and 
        Jonathan~Berger\thanks{Prateek~Verma, Jonathan~Berger are with Center for Computer Research in Music and Acoustics, Stanford University, e-mail: prateekv@ccrma.stanford.edu, e-mail: brg@ccrma.stanford.edu\newline\indent Prateek~Verma is the corresponding author.}}

\maketitle 

\begin{abstract}
% We present a framework to model and enhance perceptual, aesthetically pleasing attributes associated with audio signals by combining convolutional architectures, with ideas from classical signal processing. We transform the sound of a cheap low quality musical instrument to that of an expensive richer sounding instrument without the need of parallel data which is often hard to collect. We adapt the classical approach of a simple adaptive EQ filtering to the objective criterion learned by a neural architecture and optimize it to get the signal of our interest.  Since we learn adaptive masks depending on the signal of interest as opposed to a fixed transformation for all the inputs, we show that shallow neural architectures can achieve the desired result. A simple constraint on the objective and the initialization helps us in avoiding adversarial examples, which otherwise would have produced noisy, unintelligible audio. We believe that the current framework proposed has enormous applications, in a variety of problems where one can learn a loss function depending on the problem, using a neural architecture and optimize it after it has been learned.
We present a framework to model perceived quality of audio signals by combining convolutional architectures, with ideas from classical signal processing, and describe an approach to enhancing perceived acoustical quality. We demonstrate the approach by transforming the sound of an inexpensive musical with degraded sound quality to that of a high quality musical instrument without the need of parallel data which is often hard to collect. We adapt the classical approach of a simple adaptive EQ filtering to the objective criterion learned by a neural architecture and optimize it to get the signal of our interest.  Since we learn adaptive masks depending on the signal of interest as opposed to a fixed transformation for all the inputs, we show that shallow neural architectures can achieve the desired result. A simple constraint on the objective and the initialization helps us in avoiding adversarial examples, which otherwise would have produced noisy, unintelligible audio. We believe that the current framework proposed has enormous applications, in a variety of problems where one can learn a loss function depending on the problem, using a neural architecture and optimize it after it has been learned.
\end{abstract}
    
% \fi
\IEEEpeerreviewmaketitle

\section{Introduction}
Deep neural networks \cite{deeplearningbook} have had a profound impact across a wide range of fields including audio signal processing. It has led to rethinking fundamental approaches to classical problems such as speech recognition\unskip~\cite{339587:7515567}, Text to Speech synthesis\unskip~\cite{339587:7515071}, music generation\unskip~\cite{339587:7519520}, speech to speech translation \cite{guo2019end}, and auditory perception and cognition \unskip~\cite{339587:7515063}\unskip~\cite{339587:7515324}. Many research problems can be distilled to mapping a set of fixed/variable length vector to a single/variable length output. Due to such flexibility we have seen numerous approaches from fields like natural language processing\unskip~\cite{339587:7519508}\unskip~\cite{339587:7515267} computer vision\unskip~\cite{339587:7519509}\unskip~\cite{339587:7515324}\unskip~\cite{339587:7515065} audio and speech processing cross pollinating and in so doing yielding unique approaches across domains. 

For the current problem, of modeling human hearing perception, a lot of work in the past has been done from the biological perspective \unskip~\cite{339587:7519522}, in understanding speech in noise\unskip~\cite{339587:7519564} and also from the point of view of listening experiments in order to understand human hearing \unskip~\cite{339587:7519521}. While listening experiment in a way bring about the salience aspects of the problem of interest, we present a new way to understand and model human hearing perception via neural loss functions. For the current problem we focus on two aspects viz. how can we use a convolutional architecture to first model what constitutes the good perceptual attributes associated with audio signals for the case of instrument sounds. The approach draws inspiration from classical signal processing techniques like adaptive EQ equalization \unskip~\cite{339587:7515483} to model the weighting function applied to the time-frequency spectogram representation of audio signals. Different weights are learned  for different input signals similar which alleviates need of a fixed  high dimensional fixed mapping for all inputs. We also keep into consideration the fact that unaligned data is unavailable in most practical scenario, and there exists a strong correlation between the spectral structure of a perceptually good sounding audio and a normal audio recording. For the focus of the current work, we also understand the trained model as to what attributes of input signal makes a musical instrument sound good. 

The main contributions of this paper are :

1. We present work on modeling perceptual attributes of an audio signal using a loss function learned by a convolutional architecture, and show that we can learn which parts of the spectrogram contribute most to the desired property of interest. 

2.  We also present a way of learning a simple adaptive EQ system on the content of the signal, optimizing a learned cost function.

3.  We show a method of avoiding adversarial examples by initialization technique thus yielding meaningful spectral templates, evaluated both subjectively and empirically.

The structure of the paper is as follows: Section II and III discusses related work in vision, music, signal processing and audio perception for tackling such problems followed by the method we used to tackle the problem. The datasets used in the paper, how they were collected are presented in Section IV followed by quantitative and subjective results in Section V.  The paper is concluded by a brief discussion and interpretation of the results.

\section{Related Work}
The problem becomes challenging in a non-parallel data setup, as once the parallel data is available, standard neural mapping approaches \unskip~\cite{339587:7515268}\unskip~\cite{339587:7515269}\unskip~\cite{339587:7515067} can be tried easily and will in high-likelihood give good results. In a non-parallel setting and the recent approaches using Generative Adversarial Networks (GANs) \unskip~\cite{339587:7519608} are worthy of exploring in the future, but they are difficult to interpret them, and more like black box approaches. There do exist however similar lines of work in computer vision where perceptual characteristics are learned and enhanced when there is ground truth distributions and aligned data \unskip~\cite{339587:7515281}/unaligned data \unskip~\cite{339587:7515061}. However, there exist fundamental differences between image and audio signals with latter having periodic structure across frequency, and the representation of the time and frequency. Recently there was work done using Long-Short Term Memory (LSTMs) for denoising by directly estimating the perceptual PSEQ score \unskip~\cite{339587:7515062}  and subsequently minimizing it instead of euclidean norm between target and predicted spectogram slices\unskip~\cite{339587:7515280}. Often these network tend to overfit on synthetic additive noise samples \unskip~\cite{339587:7519817} by memorizing it.  Typically, when there is aligned data available, neural architectures have shown a lot of promise in learning to map inputs of fixed dimensions to outputs of fixed dimensions. Such mappings can be found in problems such as speech denoising, source separation, image enhancement, acoustic scene understanding and speech recognition, emotion recognition \unskip~\cite{339587:7515324}, \cite{haque2019audio},\unskip~\cite{339587:7515567}\unskip~\cite{339587:7515066}. In practice however there is not always the possibility of creating such aligned data where the input-output pairing is readily available. There are some ways of creating artificial aligned data by creating synthetic mixtures in problems such as source separation and denoising, enhancement and transcription. However, in some problems such as speech enhancement, equalization, there is usually vast amounts of unaligned data available. Traditional studies in neural network performance analysis have shown the vast amounts of data plays an important role in the performance of the system, sometimes outweighing a poor model or the number of parameters present \unskip~\cite{339587:7520159}. The current work explores a way to use such vast amounts of data.

\section{Methodology}
Inspired by recent advances in acoustic scene understanding by convolutional architectures \unskip~\cite{339587:7515324}, we begin to explore understand human hearing perception via a neural architecture. There can be a variety of attributes of an acoustic signal that can be associated with hearing perception and it depends on the type of input acoustic signal too. For the sake of this paper, we choose to model the attributes that makes cello sounds  pleasing to ears by recording a high-end expensive cello vs a normal cello. There exist several such parallel problems in different domains, and the goal of this work is to show that we can solve it for this specific task. The framework can be adapted to several such domains, where there are differentiating factors associated with two domains of sounds. In most of the cases, given the advancements of audio understanding, a convolutional architecture will also be able to differentiate the domain of interest.

An important point to note here is that we intend to retain the same spectral structure in time-frequency spectogram representations of  input and output, and change the spectral weights. Our loss function formulation and approach restricts us for the same, although with appropriate loss function tweaking it can be adapted to problems like reverberation etc. Two audio pieces recorded on different quality cellos  in similar acoustic environment, playing the same piece in the same style, and time aligned  will only differ from each other in how much weight is given to the harmonic as well as stochastic component. The primary difference between a low-end and high-end cello is how the resonances are different in both the instruments, both as a function of which notes which are being played, as well as attack and decay of individual notes.  This correlates to the strength of all the frequency bins in magnitude spectra and their overall distribution.

The approach we would follow for doing the transformation keeps this assumption in mind. In brief, we would model the attributes responsible for the perception of good sound vs better sound. We first learn a scoring network, using a convolutional architecture, which takes a log-magnitude spectogram as an input and assigns a continuous score between [0,1] with 0 being an expensive high quality cello and 1 being a cheap cello sound. Backpropagation with gradient descent, with dropout regularization is used to minimize the euclidean norm between the predicted and target scores. We achieve {\textasciitilde}93\% in distinguishing 3s of audio samples between the two classes. For all the experiments, a 512-pt FFT with a hanning window of size 30ms, hop 10ms with sampling rate of 16000Hz was used.  Mathematically $X_i $ be the input spectogram, and $X_t $ be the transformed spectogram of interest.  We define the convolutional architecture as a mapping $L_c(X) \rightarrow [0,1] $ which takes in a spectogram input and predicts a continuous number with $L_v$ computing the variance of the mask. This loss function is used to learn a spectral mask which can take an input sound and convert it to a desired sound while retaining the same spectral structure, and just weighing spectral weights at different parts of the T-F representation. The transformation is a convolution operation in the time domain which would amount to addition by a weighing function in log-magnitude domain,  

$$ X_t = X_{inp} + {M_o} $$
$$ M_o = argmin_{M} L_c(X_{inp}+{M}) + \alpha ||X_{inp}-X_t||^{2} + \beta L_v({M}) $$ 

 Notice how the approach looks a bit similar to having some of the GANs,\unskip~\cite{339587:7519608}  but a close inspection reveals the differences We learn a separate mask, depending on the type of input, and optimize it as opposed to having a high dimensional generic mapping for all the inputs \unskip~\cite{339587:7531687}. This allows us to have a very shallow, small number of parameters instead of a complex architecture, and direct similarities with classical techniques. We also do not alternate between training the masking function and the scoring function. The scoring function is trained first, by  tuning the model sizes to achieve good performance on a held out test set with as few parameters as possible. This network will then be used as a scoring function to learn adaptive masks. We borrow some of the technique from traditional signal processing algorithms like binary masking and learning adaptive filters.  There has been some work on content adaptive neural architectures too where it has been shown that a smaller model can achieve comparable accuracy to a fixed model with large number of parameters \unskip~\cite{339587:7531687}. The masking strategy can potentially suppress noise, change the weights of the input spectogram while preserving the spectral template, adjust the strength of the transients etc.  

\textbf{Avoiding adversarial examples}: The salient point of our approach would also be how we circumvented adversarial examples by incorporating various constraints into our loss functions and type of initialization. One of the biggest weakness of neural architectures is the ability of being vulnerable to adversarial examples. This is a well studied topic with such inputs being present in vision\unskip~\cite{339587:7515271}, speech and natural language processing\unskip~\cite{339587:7515272}. Small amounts of imperceptible noise can be added to the input signal to make the classifier predict the category of our interest. The initial experiments in our method also resulted in learning a template function which had similar noisy, unintelligible structure, while minimizing the scoring function learned to score the quality of two audio samples as seen in Figure~\ref{figure-06b49f8a86a695e52cbeaaa9dfe819fd}.

\bgroup
\fixFloatSize{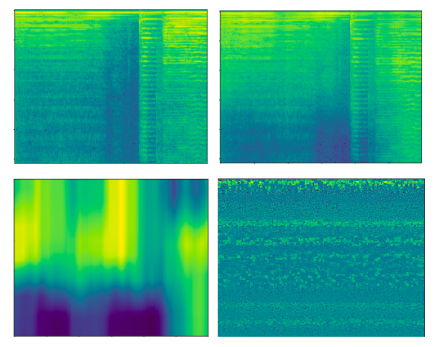}
\begin{figure}[!htbp]
\centering \makeatletter\IfFileExists{503cce63-7b64-416b-8e7d-5a96d051b8ff-uscreen-shot-2018-11-01-at-2-u38-u59-pm.png}{\includegraphics{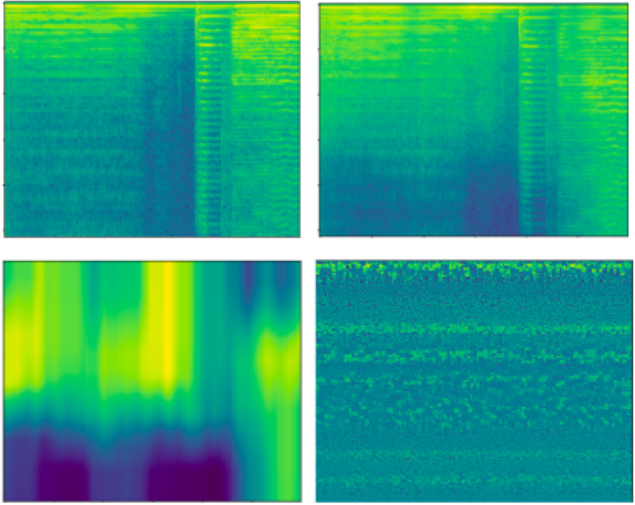}}{}
\makeatother 
\caption{{Top Left: Original input Top Right: Transformed input to an expensive cello Bottom Left: Mask learned with rectangular block initialization Bottom Right: Mask learned with random initialization resulting in a noisy adversarial output.}}
\label{figure-06b49f8a86a695e52cbeaaa9dfe819fd}
\end{figure}
\egroup
Figure~\ref{figure-06b49f8a86a695e52cbeaaa9dfe819fd} shows the transformation we intended to make, and the noisy mask which we learned with having alpha and beta as 0. In order to avoid the problem, we pose additional constraints on the learning mask. We know that the output spectogram would have similar template as the input spectogram, and the weights would not change drastically from one time frequency bin to the next. So the objective criterion incorporated additional constraints using the euclidean distance of the input spectogram and the target spectogram. Additionally, we initialize the input mask with large blocks instead of random Gaussian noise to every pixel. This assumption is again due to the structure of audio spectograms and the transformation we intended to make. The initial blocks are then averaged to yield a smoother transitions of the weights of the mask.

\bgroup
\fixFloatSize{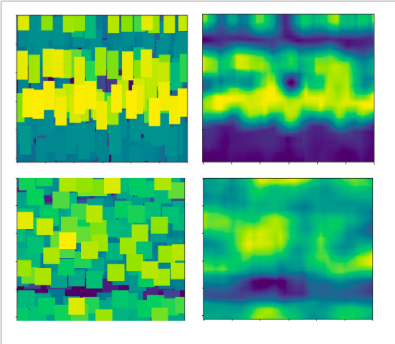}
\begin{figure}[!htbp]
\centering \makeatletter\IfFileExists{afddd1e3-00a6-41b3-bb3b-b364d1019438-uscreen-shot-2018-11-01-at-12-u29-u47-pm.png}{\includegraphics{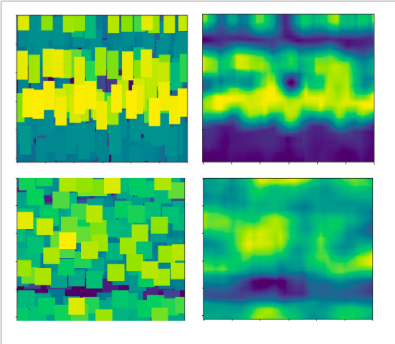}}{}
\makeatother 
\caption{{Final optimal rectangular and square block weights of the mask of two different audio samples on left before smoothing and the smoothed version on the right. }}
\label{figure-2fc269fb1c1a80c02127a8601faae9e2}
\end{figure}
\egroup
Figure~\ref{figure-2fc269fb1c1a80c02127a8601faae9e2} shows how we can tweak the surface and gain of the time-frequency weighing function by the width, and height of the initialization blocks as well as number of such blocks, for two different audio pieces. (Naively the initialization strategy is analogous to putting sticky notes on the spectogram where we could control the position, size, shape, number and their gains to create a mask to be added to the spectogra)

\section{Datasets}
For our problem, since there was no publically available dataset available, we decided to create one by fixing up the instrument to be cello. A point to note is that the choice can work for  instrument/speech pairs for almost all the cases. For majority of cases where we want to transform the perceptual attributes, the spectral template remains the same and the only thing that is changing is the weight assigned to the time-frequency representation. 

Roughly 5 hours of recordings were made using a high end cello (costing {\textasciitilde} \$150,000) which musicians preferred to play over a regular cheaper cello. Initial experiments revealed that the network was picking up on the impulse response instead of the perceptual attributes as a differentiator, so we decided to fix up different attributes such as pieces, player, microphone, acoustic environment same except for the difference a normal vs high end cello. The musician always preferred to play on the expensive cello due to sheer beauty and the richness of the sound. YouTube recordings were not taken as there arises differences in reverberation which the current architectures would not be able to tackle. However, adding a model which can dereverb an audio signal can be easily adapted to the current framework without having any change in the objective criterion or the objective criterion can be modified for such cases. For evaluation experiments the audio pieces were first manually coarse aligned and then aligned using dynamic time warping to take into account the differences in onset times and note duration.

\section{Discussion}
In order to better understand what the neural architecture learns, we used a popular technique in computer vision which creates saliency map to highlight portions of the image contributing most towards a particular class for audio spectograms\unskip~\cite{339587:7515064}. Popularly known as class activation maps, for a convolutional architecture, it yields to better interpretability of these high dimensional black-boxes and often provide meaningful insights in understanding these systems.

\bgroup
\fixFloatSize{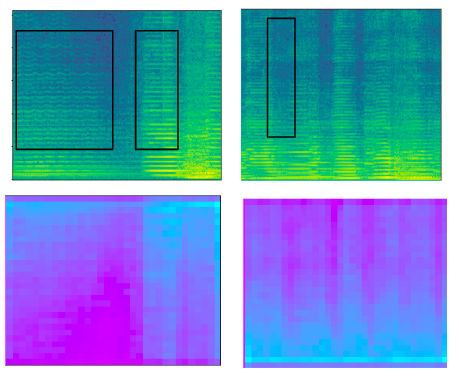}
\begin{figure}[!htbp]
\centering \makeatletter\IfFileExists{9879408f-a65b-45f2-8726-a1ef44e1810a-uscreen-shot-2018-10-18-at-2-u50-u09-pm.png}{\includegraphics{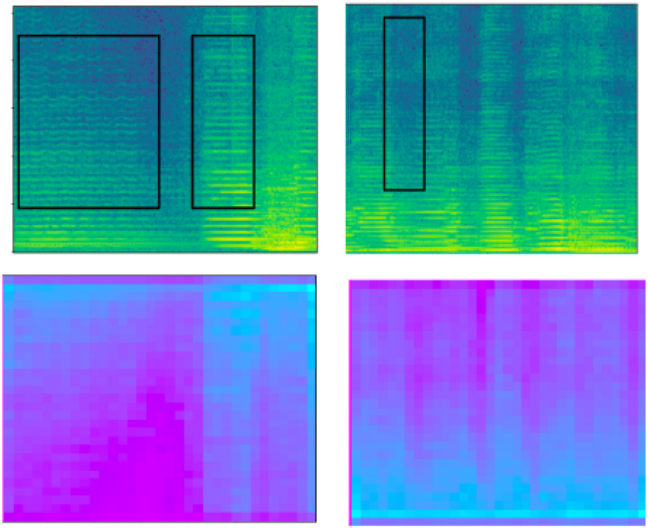}}{}
\makeatother 
\caption{{Two spectograms of expensive cello sounds with spectogram on top and the saliency maps on the bottom. Note that blue corresponds to smaller values and red corresponds to larger weights. Highlighted region depicts the focus on vibrato, transitions more than the steady regions. }}
\label{figure-303ddd282c8e9343c4077df6a0351c44}
\end{figure}
\egroup
Figure~\ref{figure-303ddd282c8e9343c4077df6a0351c44}  shows two such maps for an expensive cello. There are a couple of points which are fascinating here in observing these saliency maps. We see that in order to characterize an expensive high-end cello, the net gives more weightage to the vibrato portion of the spectogram than the steady regions. The same is observed for the stochastic regions of the signal as well. Even though both of the pieces were recorded in the same acoustic environment, the manner in which the note changes, plays a crucial role in determination of how good a cello is. This was confirmed by an experienced cellist too, that listening to the same piece being played both a poor and a high-end cello, these differences were felt. During close inspection of masks learned from rectangular tile initialization, in  Figure~\ref{figure-06b49f8a86a695e52cbeaaa9dfe819fd}  we see that the optimal mask also gives more weight to the stochastic regions on a chosen audio consisting of such regions. The observation that our mask can change according to the type of input signal and  adapt itself to enhance/suppress certain regions of the spectograms is also worth noting.

\section{Conclusion and Future Work}
This paper has explored a way to model perceptual qualities of human hearing  and from classical signal processing, derived a novel technique of learning adaptive weighted masking in order to enhance it. The architecture does not rely on hand coded features, and is learned end to end depending on the problem and input of interest. We also adapt the weights dynamically according to the audio signal as opposed to learning high dimensional fixed mappings. The loss learned is used to enhance the input and adapt it according to the learned perceptual metric. We see that we could achieve the desired transformations, and potential quality improvements even when aligned data was not available to us, with similar results. By incorporating constraints on temporal and spectral content as well as prior constraints of the mask, we avoid the noisy adversarial solutions. It will be interesting to try some of the recent works using adversarial architectures such as CycleGAN which also do not require unaligned data set too, although it would learn a generic mapping for all inputs. This work has a lot of potential in similar applications where we can model the perceptual  attributes or any desired characteristics optimize it using appropriate loss functions, such as text to speech synthesis, denoising, music synthesis etc.


\section*{Acknowledgment}The authors would like to thank Chris Chafe for valuable insights into cello sounds, and for help in making the recordings. We are grateful to Stephen Harrison, Department of Music at Stanford for lending us an expensive rare 17th century cello, and Stanford Artificial Intelligence Laboratory for the use of their computing resources.



%\bibliographystyle{IEEEtran}

\bibliographystyle{IEEEtran}

\bibliography{article}

\end{document}